\documentclass[aps,prl,twocolumn,showpacs,superscriptaddress,groupedaddress, showkeys]{revtex4}  
\usepackage[dvips]{graphicx}

\usepackage[]{amsfonts}
\usepackage[]{amssymb}
\usepackage[]{bm}
\usepackage[]{amsmath}

\begin{document}

	\title{A simple electronic device to experiment with the Hopf bifurcation}

        \author{Daniela N. Rim}
	\affiliation{Faculty of Exact and Natural Sciences, National University of Cuyo. Padre Contreras 1300, 5500 Mendoza, Argentina.} 
	
	\author{Pablo Cremades}
	\affiliation{Faculty of Exact and Natural Sciences, National University of Cuyo. Padre Contreras 1300, 5500 Mendoza, Argentina.}
	
	\author{Pablo Kaluza}
	\affiliation{Interdisciplinary Institute of Basic Sciences, National Scientific and Technical Research Council (CONICET) \& Faculty of Exact and Natural Sciences, National University of Cuyo. Padre Contreras 1300, 5500 Mendoza, Argentina.} 
	

	\date{\today}

	\begin{abstract}We present a simple low-cost electronic circuit that is able to show two different dynamical regimens with oscillations of voltages and with constant values of them. This device is designed as a negative feedback three-node network inspired in the genetic repressilator. The circuit's behavior is modeled by a system of differential equations which is studied in several different ways by applying the dynamical system formalism, making numerical simulations and constructing and measuring it experimentally. We find that the most important characteristics of the Hopf bifurcation can be found and controlled. Particularly, a resistor value plays the role of the bifurcation parameter, which can be easily varied experimentally. As a result, this system can be employed to introduce many aspects of a research in a real physical system and it enables us to study one of the most important kinds of bifurcation.
	\end{abstract}

	
	
	
	\maketitle

	
	\section{Introduction}
	\label{sec_introduction}

	Physical systems, generally, are mathematically modeled by differential equations and studied in the framework of the dynamical system theory \cite{Strogatz-book,Perko}. This formalism is also applied to many other systems like biological and economic ones \cite{Murray}. The main goal in this theory is to find the behavior of the solutions or the trajectories in the space of system variables and, in particular, how they evolve to different kind of attractors, like fixed points and limit cycles. Notably, many systems have the property of having several qualitatively different solutions as a function of a certain parameter. The bifurcation analysis is the part of dynamical systems that studies the way in which these changes of dynamics arise with respect to these bifurcation parameters. 
    
        The Hopf bifurcation is one of the most important and known of the bifurcations in this theory. It is characterized by changes of dynamics between stationary states (stable fixed points) with fixed values for the system variables, and, oscillatory dynamics (limit cycles) where these variables evolve periodically in time. The system dynamics are controlled by a bifurcation parameter which is in general an important and characteristic parameter of the system. 
    
        This bifurcation emerges in many mathematical models that intent to describe different real systems. For example, the Hodgkin-Huxley \cite{HH-model} and the FitzHugh-Nagumo  \cite{FN-model} models for the neural membrane potential. Non-linear chemical oscillators like the Belousov-Zhabotinsky reaction \cite{BZ-model} with the Oreganator \cite{oreganator-model} model as a possible mechanism of operation, and, the Brusselator \cite{brusselator-model}. Some special cases of the predator-prey model \cite{PP-model} for population dynamics. The Lorentz attractor \cite{lorentz-model} as a paradigmatic example of deterministic chaos. And finally, we mention the repressilator \cite{repressilator-model} as an example of a genetic network.
    
        We can figure out from the previous paragraph that the Hopf bifurcation plays an important role in the characterization of many systems. However, most of the systems that the previous models intent to characterized are quite difficult to implement in a laboratory for undergraduate students.  We propose in this work to design a very simple electronic circuit that is able to behave as having a Hopf bifurcation. We do not only observe typical oscillations, but we can also control the bifurcation parameter (resistor) in order to observe all the regimens of this bifurcation.
    
        The proposed circuit is based on the genetic repressilator \cite{repressilator-model}. It is an artificial genetic regulatory network that consists of three genes that repress each other in a loop structure. It is a paradigmatic example of the \textit{synthetic biology} field \cite{synthetic-biology}. Although very sophisticated electronic devices have been proposed for genetic systems \cite{paper-kurst}, the main principle of operation of the repressilator can be mimicked with a simple $RC$ circuit. In effect, some of these circuits have been implemented in order to study synchronization properties on networks of artificial like-genes \cite{japoneses-circuitos}.
    
        This manuscript is organized as follows. In the next section, we introduce the electronic circuit and the mathematical model to describe it. We also develop a linear stability analysis and we find the eigenvalues of the stability matrix. In the third section, we present the analytical, numerical and the experimental results. Finally, we discuss this work and present the conclusion in the last section.

	\section{Model}
	
\begin{figure}[!ht] 
            \begin{center}
                \includegraphics[width=1.0\columnwidth, angle =0, clip]{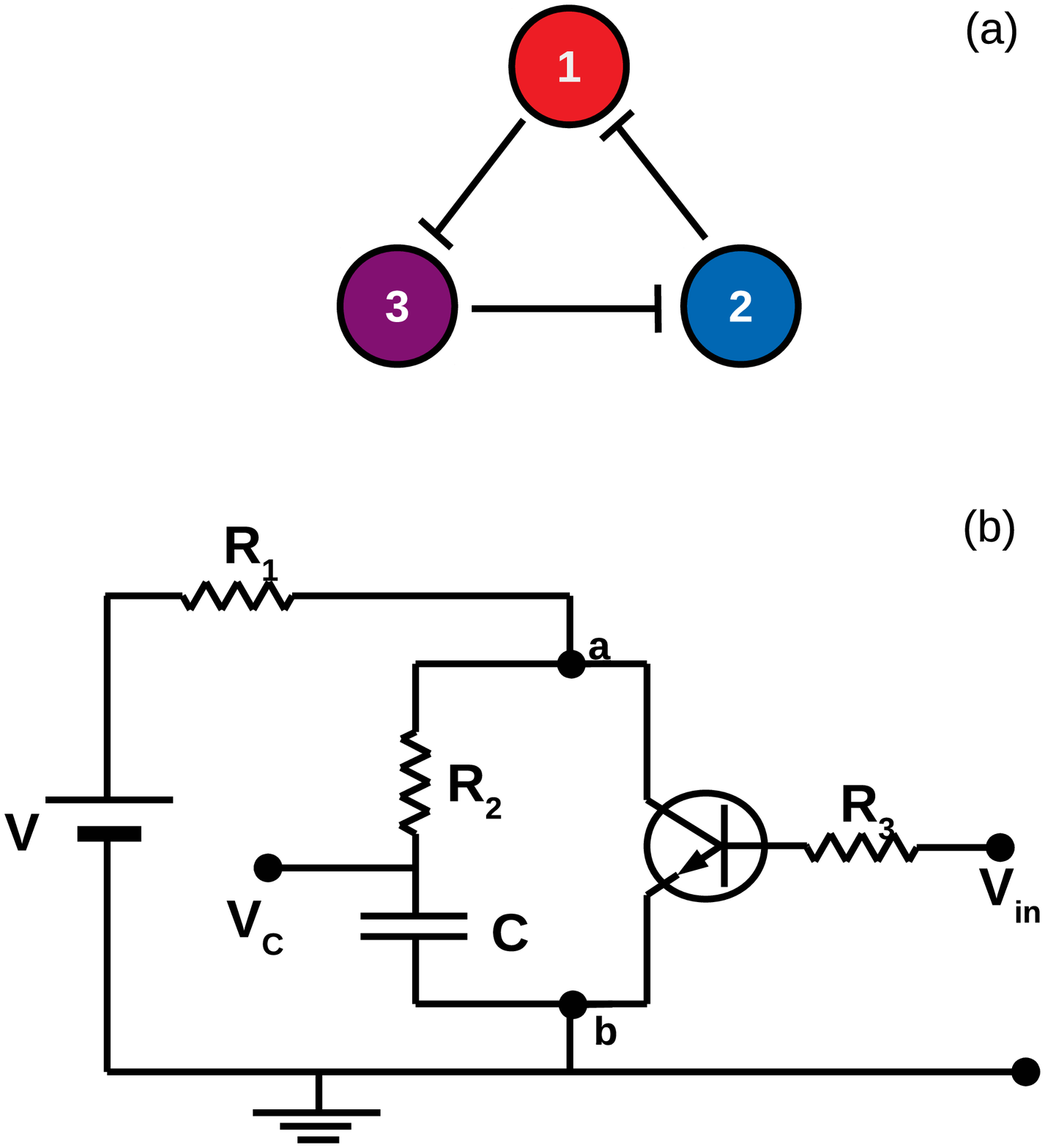}
                \caption{(a) Schematic representation of a repressilator-like system. (b) Electronic circuit of a node. The node is characterized by the output signal $V_C$ and it is controlled by the input signal $V_{in}$.} 
                \label{fig_circuito_node} 
            \end{center}
        \end{figure}

	The genetic repressilator consists of three genes that repress each other in a loop-like structure as it is shown in Fig. \ref{fig_circuito_node}(a). The flat arrows are the biological representation of a repressive directed interaction from a node to the other. We can give an insight of the operation of this system for certain parameter values as follows. Each node or gene is characterized to behave as a bistable unit with high and low levels of expressions. Thus, when a node has a high expression it represses strongly the node that it points to. The effect in the second node is to reduce its expression and takes the lower level. And so it results in a reduction of the repression of the third node that takes a high state. It is clear that this chain of effects cannot be stabilized since the system presents a kind of \textit{frustration} and oscillatory dynamics emerge.

        \subsection{Circuit design}

        As a first task, we designed an electronic node that is able to present dynamics with two levels of expression.  A node in this circuit can be seen in Fig. \ref{fig_circuito_node}(b). This is a simple $RC$ circuit where the path of charge and discharge of the capacitor can vary depending on the transistor state which operates basically as a switch. The expression level of this device is given by $V_C$, and, the input signal is given by $V_{in}$.
    
        In the first case, when $V_{in} = 0$, the transistor is in the \textit{cut-off region} and does not allow current to flow between its collector and emitter, and the circuit is described as
    
        \begin{equation}
            \frac{dV_{C}}{dt} = \frac{V - V_C}{(R_1 + R_2)C}.
            \label{equ_cut_off}
        \end{equation}
    
        \noindent
        Here, the capacitor will get a charge $V_C = V$ with a time constat $\tau_1 = (R_1+R_2)C$. 
    
        In the second case, when $V_{in}/R_{3} \gg h$, the transistor is in the \textit{saturation region} and it conducts current between the collector and the emitter, and the system is described by
    
        \begin{equation}
            \frac{dV_{C}}{dt} = - \frac{V_C}{R_2C}.
            \label{equ_saturation}
        \end{equation}
    
        \noindent
        In this case the capacitor discharges to zero with a time constant $\tau_2 = R_2C$. The parameter $h$ indicates the minimum value of current needed in order to activate the transistor.
    
        We note that when current $V_{in}/R_3$ in the base of the transistor is not enough to reach the saturation region the circuit behaves as a combination of the two previous cases. We can model this general situation like 
    
        \begin{equation}
            \frac{dV_{C}}{dt} = \frac{1}{\bigg(R_1u(V_{in}) + R_2\bigg)C}\bigg( Vu(V_{in}) - V_C\bigg).
                \label{equ_un_nodo}
        \end{equation}
    
        \noindent
        In this equation the step-like function 
    
        \begin{equation}
            u(x) = \frac{1}{2}\Bigg[\tanh \Bigg( \beta \bigg( h - \frac{x}{R_3} \bigg) \Bigg)  + 1 \Bigg].
        \end{equation}
    
        \noindent
        The function $u(x)$ is an approximated description of the behavior of this circuit with a \textit{NPN} transistor. Parameters $\beta$ and $h$ depend on the internal properties of the transistor and the other elements of the circuit. $\beta$ is the current gain of the transistor defined as:
        
        \begin{equation}
            \beta = I_b/I_c
        \end{equation}

        \noindent
	where $I_b$ is the base current and $I_c$ is the collector current. $h$ is a hybrid parameter of the transistor in common emitter configuration. It depends on the quiescent point of the transistor, i.e., the external components in the circuit. As a result of this design, the voltage $V_C$ can vary between $V$ and zero depending on $V_{in}$. In particular, $V_{in}$ plays the role of a repressive signal since $V_C$ behaves vice verse to $V_{in}$.

        In the next step, we connect three of these nodes in a loop in order to construct a negative feedback closed-loop network. The output voltage $V_i$ of a node (previous $V_C$) is used as input ($V_{in}$) of the next node. The dynamics for such a system is as follows
    
        \begin{eqnarray}
            \frac{dV_{i}}{dt} &=& F_{ij}(V_i, V_j) = \nonumber \\ 
                      & & \frac{1}{\bigg(R_1u(V_j) + R_2\bigg)C}\bigg( Vu(V_j) - V_i\bigg).
                \label{equ_repressilator}
        \end{eqnarray}
    
        \noindent
        Here, only the elements $F_{12}(V_1,V_2)$, $F_{23}(V_2,V_3)$ and $F_{31}(V_3,V_1)$ are different from zero. 
    
        The model we obtain in (\ref{equ_repressilator}) is a non-linear system of coupled first order differential equations. The non-linearity is the result of function $u(V)$ with its sigmoid shape ($\tanh()$). Additionally, as a result of this particular function, we cannot find a closed-form expression for the solution to this system. Note that this kind of expressions are common for modeling circuits with transistors. For example, the work cited in ref. \cite{japoneses-circuitos} also employs a similar description for its device.

        \subsection{Stability analysis}
    
        We start this analysis of the mathematical model by searching the fixed points of the system (\ref{equ_repressilator}), and later, we develop the linear stability analysis on this point.
    
        \subsubsection{Fixed point}
        
        The fixed point of the system (\ref{equ_repressilator}) is the point in the space of voltages $\{V_i\}$ ($i = 1,2,3$) where the condition $\frac{dV_i}{dt}=0$ is satisfied. In order to find this point we argue in the following way. The three coupled equations are identical among them with a strong symmetry with respect to the voltages (circular reference). As a result, they must fulfill the same condition simultaneously. That is:
        
        \begin{equation}
            u(V_p) = \frac{V_p}{V}.
        \end{equation}
        
        \noindent
        Here, $V_p$ is the voltage that the capacitors must have in order to settle the system on the the fixed point. Although we cannot find a close-form expression for $V_p$ we can observe that it corresponds to the intersection of a straight ($y = V_p/V$) and a negative hyperbolic tangent ($y = -tanh(V_p)$). Thus, it is simple to see that there is only one possible intersection of these two curves. As a result the system has always only one fixed point. We call this point $\vec{P} = (V_p,V_p,V_p)$.

    
    
        
        Since we cannot find explicitly the value of $V_p$ because the characteristics of the function $u(x)$, we propose to get this value numerically by reducing the quantity 
    
        \begin{equation}
            \epsilon(V_p) = \Bigg|  u(V_p) - \frac{V_p}{V} \Bigg|.
            \label{equ_error}
        \end{equation}
    
        \noindent
        This miminization is performed by variying $V_p$ between zero and $V$ each small intervals $\Delta V_p$ and computing the quantity $\epsilon(V_p)$. A small value of $\Delta V_p$ ensures a good aproximation for the position of the fixed point.

        \subsubsection{Linear stability analysis}
    
        We evaluate the Jacobian matrix of the system (\ref{equ_repressilator}) on the fixed point $\vec{P}$. The elements of this matrix are    
        
        \begin{equation}
            A = \frac{\partial F_{ij}}{dV_i} \Bigg|_{\vec{P}} = -\frac{1}{\big(R_1u(V_j) +R_2 \big)C} \Bigg|_{\vec{P}} 
        \end{equation}
        
        \noindent
        and
        
        \begin{eqnarray}
        B = \frac{\partial F_{ij}}{dV_j}\Bigg|_{\vec{P}} &=& \frac{u'(V_j)}{C} \Bigg[ \frac{V}{(R_2 + R_1u(V_j))} \\ \nonumber
                                                            &-& \frac{V_iR_1 + VR_2}{\big(R_1u(V_j) +R_2 \big)^2} \Bigg] \Bigg|_{\vec{P}}, 
        \end{eqnarray}
    
        \noindent
        with
        \begin{equation}
            u'(x) = -\frac{1}{2}\frac{\beta}{R_3} sech^2 \Bigg( \beta \bigg( h - \frac{x}{R_3}\bigg) \Bigg).
            \label{equ_funcion_du}
        \end{equation}
    
        The Jacobian matrix has the shape
    
        \begin{equation}
            J = \begin{bmatrix}
                    A & 0 & B \\
                    B & A & 0 \\
                    0 & B & A
            \end{bmatrix}.
            \label{equ_jacobiano}
        \end{equation}
    
        \noindent
        This matrix has the following eigenvalues
        
        \begin{eqnarray}
            \lambda_1 &=& A + B, \\ \nonumber
            \lambda_2 &=& A - \frac{1}{2}B + \frac{\sqrt{3}}{2}B\bold{i}, \\ \nonumber
            \lambda_3 &=& A - \frac{1}{2}B - \frac{\sqrt{3}}{2}B\bold{i}.
            \label{equ_autovalores}
        \end{eqnarray}
    
        In order to know the kind of dynamics as a function of $R_3$, we need to find the sign of the real parts of these eigenvalues. In particular, we need to know if $Re(\lambda_{2,3})=0$ for a critical value $R_c$ for the bifurcation parameter $R_3$. Unfortunately, we cannot find an analytical formula for this dependence and we need to evaluate these expressions numerically.

        \section{Results}
    
        In this section, we present several results of this device and its mathematical model. We show numerical simulations and experimental work.

        \subsection{Experimental study}
    
\begin{figure}[!ht] 
            \begin{center}
                \includegraphics[width=1.0\columnwidth, angle =0, clip]{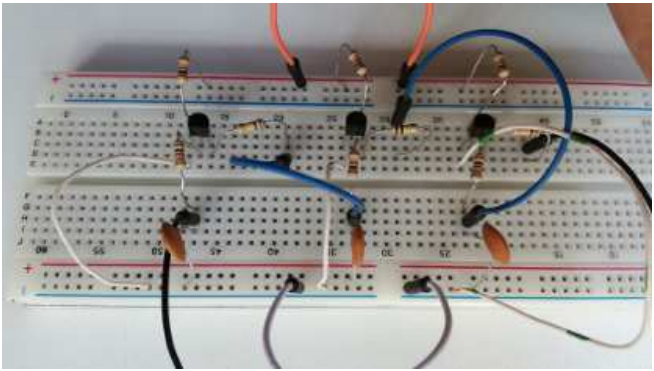}
                \caption{Experimental setup implemented on an experimetal breadboard.} 
                \label{fig_circuito_foto}
            \end{center}
        \end{figure}

        We constructed the experimental device whose design is shown in Fig. \ref{fig_circuito_node}(b). The values of the electronic elements are: $R_1 = 1 K\Omega$, $R_2 = 1 K\Omega$, $C = 220nF$, $V = 5.45V$. The transistor is type $NPN$  model $2sc2235$. The experimental setup is shown in Fig. \ref{fig_circuito_foto}.  We note that in the experimental device there are variations on the values of the electronic components between nodes because of the manufacturing dispersion. As a result, the analytical derivations we have performed before are qualitatively correct if these variations are relatively small.
    
        \begin{figure}[!ht] 
            \begin{center}
                \includegraphics[width=1.0\columnwidth, angle =0, clip]{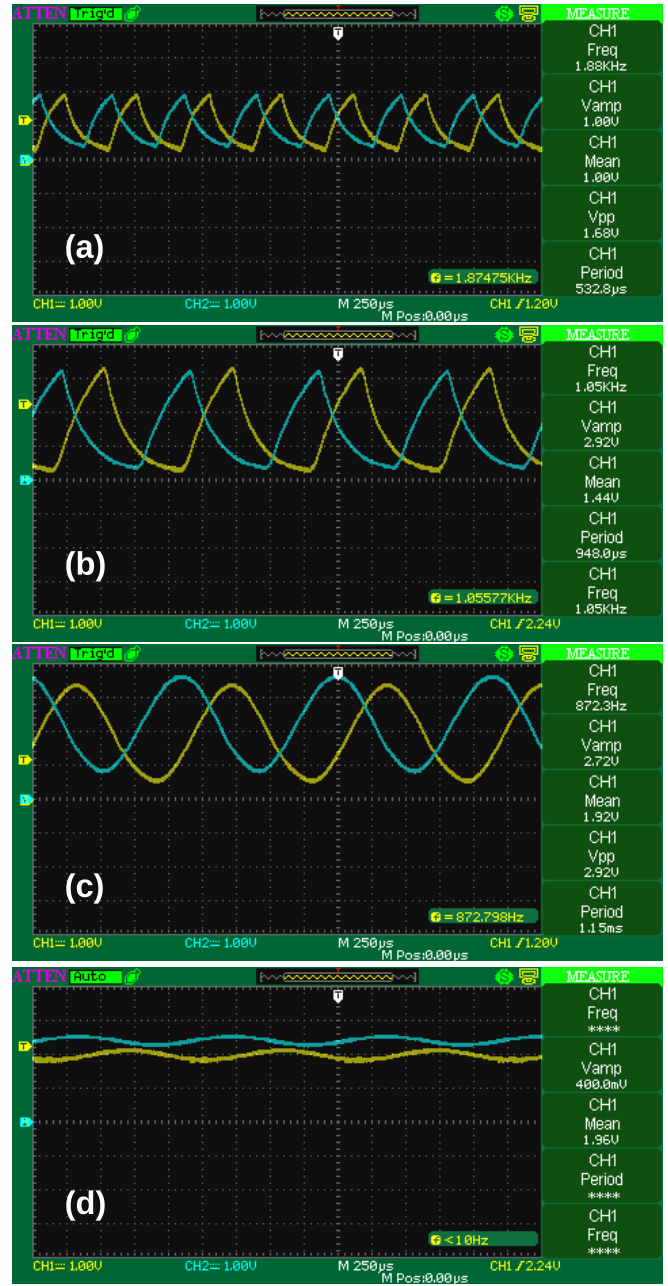}
                \caption{Oscilloscope measurements of voltages $V_i$ as a function of time after the transients for two nodes. In each figure the control parameter has different values: (a) $R_3 = 1 k\Omega$, (b) $R_3= 47 k\Omega$, (c) $R_3 = 100 k\Omega$ and (d) $R_3 = 110 k \Omega$.} 
                \label{fig_osciloscopio} 
            \end{center}
        \end{figure}

        We measure the voltages $V_i$ in the capacitors using an oscilloscope as a function of different values of $R_3$. The initial condition of the device is discharged with the power supply off. After turning on the voltage source $V=5.45V$ and after a transient the output voltages $V_i$ oscillate or reach constant values in time. Fig. \ref{fig_osciloscopio} presents four measurements for different values of $R_3$ where oscillations are established. We find that the frequencies are higher for smaller values of $R_3$ and that oscillations cannot be found after a critical value $R_c = 110 K \Omega$ for $R_3$ (Fig. \ref{fig_osciloscopio}(d)). Table \ref{tabla} summarizes the set of measurements for this device.

        \begin{table}[!ht]
            \begin{center}
                \begin{tabular}{ | c | c | c  | c | c |}
                \hline
                $R_3$ [K$\Omega$] & Frequency [KHz] \\ 
                \hline
                1 & 1.88 \\ 
                2.2 & 1.38 \\ 
                5 & 1.14 \\ 
                10 & 1.11 \\ 
                20  & 1.08 \\
                47  & 1.07  \\
                100  &  0.872\\
                \hline 
            \end{tabular}
            \caption{Frequencies as a function of $R_3$ for the experiemntal device.}
            \label{tabla}
            \end{center}
        \end{table}


        We observe in Fig. \ref{fig_osciloscopio}(a) that the voltages of the capacitors possess an exponential charge and discharge that can be easily seen. In effect, with a frequency of $1.88KHz$ the period $T=5.32 \times 10^{-4} s$ is around the addition of the two characteristic times of this system: $\tau_1 = R_2C = 2.2 \times 10^{-4}s$ and $\tau_2 = (R_1 + R_2)C = 4.4 \times 10^{-4}s$. On the other hand, when the system is close to the bifurcation point (Fig. \ref{fig_osciloscopio}(c) with $R_3=100K\Omega$) the signals are almost sinusoidal as we expect from the theory of Hopf bifurcations.
    
        Finally, we have measure the three output voltages when oscillations are absent for $R_3 \gg R_c$ ($R_3 = 200 K \Omega$). We find that the voltages are: $V_1 =2.5V$, $V_2  = 2.93V$ and $V_3  = 2.96V$. Note that the three voltages are different, and this differs from the theoretical analysis from which we expect the same values. This variations in the fixed point position are in part due to the dispersion of values of the electronic components.

	\subsection{Numerical analysis}
	
	In this subsection we present numerical studies in order to compute the voltage values on the fixed point and the eigenvalues of the stability matrix. We also share some examples of the possible dynamics that this model can show. In order to have compatible results in our simulations with the experiment, we chose $\beta$ and $h$ values that make fit both behaviors in the region where the Hopf bifurcation take place. As a result we take $\beta = 8.5\times10^4$ and $h=1.8\times 10^{-5}$. 
    
\begin{figure}[!ht] 
            \begin{center}
                \includegraphics[width=1.0\columnwidth, angle =0, clip]{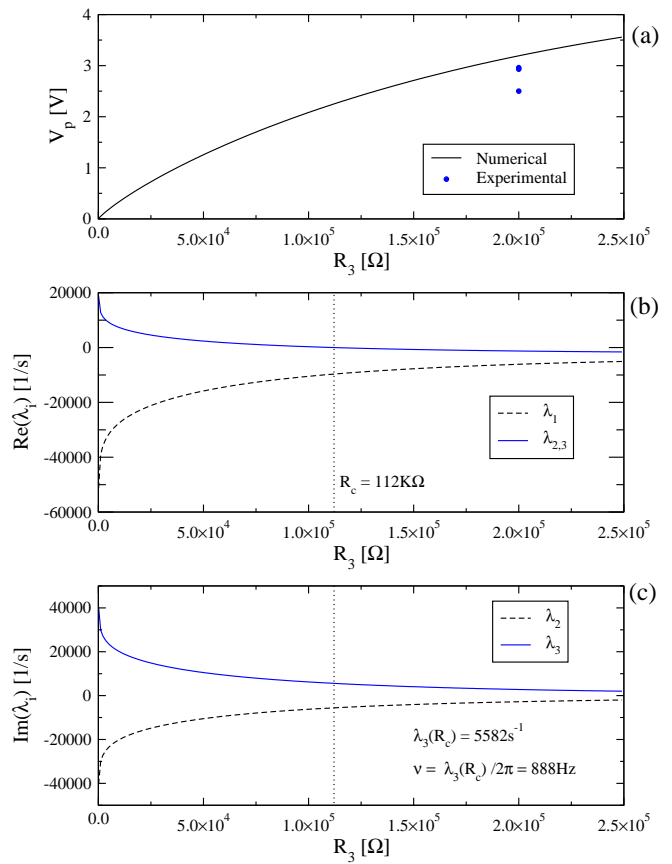}
                \caption{Fixed point and eigenvalues. (a) Value of $V_p$ as a function of the bifurcation parameter $R_3$ and experimental values (blue dots) for $R_3=200 K \Omega$. Real parts of the eigenvalues (b) and imaginary parts of the eigenvalues (c).} 
                \label{fig_puntos_autovalores} 
            \end{center}
        \end{figure}

        Fig. \ref{fig_puntos_autovalores}(a) presents the position of the fixed point $V_p$ as a function of the bifurcation parameter $R_3$. The solid curve is calculated numerically by reducing the quantity $\epsilon(V_p)$ (eq. \ref{equ_error}). We also include the experimental values obtained for $R_3=200 K\Omega$ (blue dots). We observe that these points are similar qualitatively to the proposed model.
    
        Fig. \ref{fig_puntos_autovalores}(b) and \ref{fig_puntos_autovalores}(c) present the real and imaginary parts of the three eigenvalues of the Jacobian matrix as a function of $R_3$. We find that $\lambda_1$ is real and always negative indicating that the fixed point is stable in its direction. The eigenvalues $\lambda_2$ and $\lambda_3$ are complex conjugates. Their $Re(\lambda_{2,3})$ is positive for values smaller than a critical  value $R_c = 112 K \Omega$ and negative for larger ones. This indicates that the fixed point is unstable for $R_3 < R_c$ and stable for $R_3 > R_c$. In $R_3 = R_c$ we have the Hopf bifurcation point. In this point $|Im(\lambda_{2,3}(R_c))| = 5582$, thus, the frequency of the oscillations close to the onset of the bifurcation is $888 Hz$. Note that with our election of $\beta$ and $h$ we can fit the main properties of the Hopf bifurcations with our model.

        \begin{figure}[!ht] 
            \begin{center}
                \includegraphics[width=1.0\columnwidth, angle =0, clip]{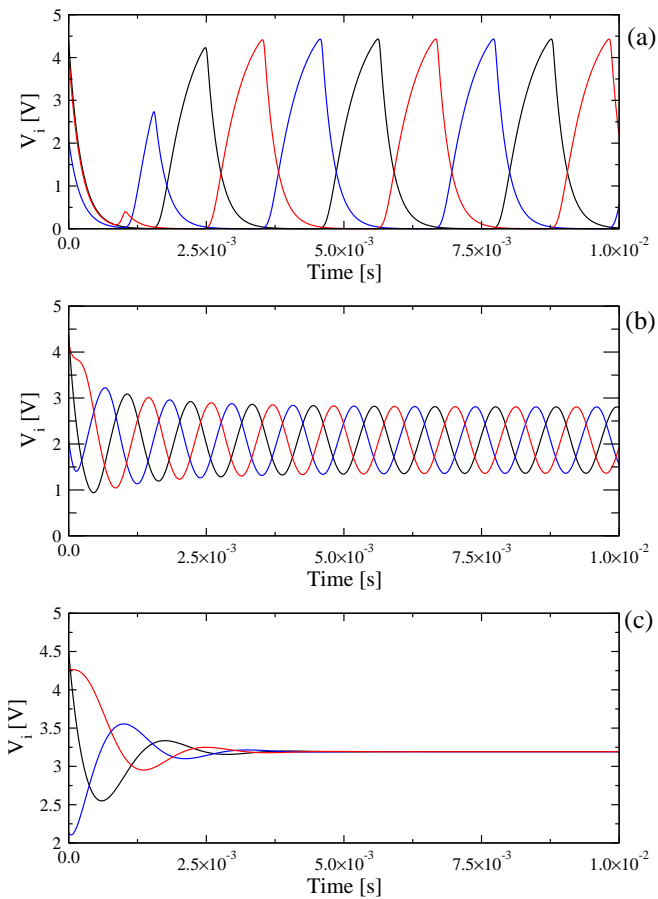}
                \caption{Voltages $\{V_i\}$ as a function of time after a transient for three characteristic values of the bifurcation parameter: $R_3=1 K \Omega$ (a), $R_3 = 100 K \Omega$ (b) and $R_3 = 200 K \Omega$ (c).} 
                \label{fig_trajectories} 
            \end{center}
        \end{figure}
    
        \begin{figure}[!ht] 
            \begin{center}
                \includegraphics[width=1.0\columnwidth, angle = 0, clip]{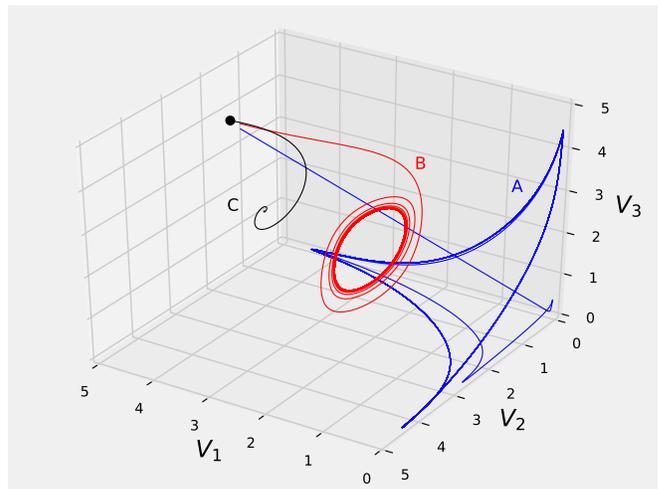}
                \caption{Trajectories from fig. \ref{fig_trajectories} in the voltage space $\{V_i\}$. The three trajectories start from the black dot. Trajectory $A$ (blue) corresponds to $R_3 = 1 K \Omega$ (fig. \ref{fig_trajectories}(a)), trajectory $B$ (red) corresponds to $R_3 = 100 K \Omega$ (fig. \ref{fig_trajectories}(b)) and trajectory $C$ corresponds to $R_3 = 200 K \Omega$ (fig. \ref{fig_trajectories}(c)).} 
                \label{fig_3d_trajectories} 
            \end{center}
        \end{figure}

        We present some characteristic trajectories in Fig. \ref{fig_trajectories}. They are performed by integrating the system (\ref{equ_repressilator}) with a Runge-Kutta algorithm of second order and a time step $\Delta t = 1\times 10^{-5} s$.  In Fig. \ref{fig_trajectories}(a) we show an evolution far from the onset of the bifurcation ($R_3 = 1 K \Omega$). We observe that the voltages show that the capacitors charge and discharge almost completely, generating signals with big amplitudes. The frequency of this signals is $317 Hz$. Note that these results are quite different from our experiment where the device has a frequency of $1.88 K Hz$ and smaller amplitude in the signals. These differences arise because our model does not represent well the transistor in the linear region of operation. The transition between cut off and saturation region is determined by the  resistor $R_1$. In this circuit the transistor enters to the active region before the capacitor completely discharges.
    
        Fig. \ref{fig_trajectories}(b) presents the voltages as a function of time close to the bifurcation point with $R_3 = 100 K \Omega$. We observe that the signals have a sinusoidal shape and relatively small amplitude. These two characteristics are typical of the Hopf bifurcation. The frequency of these signals is approximately $910 Hz$ which are in the order of the expected value. In Fig. \ref{fig_trajectories}(c) we show the evolution for $R_3 = 200K \Omega$ where the fixed point is stable. We observe that all the signals decay with oscillations to a constant value.
    
        Finally, in Fig. \ref{fig_3d_trajectories} we present the three previous trajectories in the space of voltages $\{V_i\}$. All the trajectories start from the same point and they evolve in quite different ways. When $R_3 = 200 K \Omega$ the trajectory is a spiral going to the fixed point (black curve). Close to the bifurcation point ($R_3=100K \Omega$) the trajectory evolves to a limit cycle (red curve) with an almost circle shape on a plane. In the last case, with $R_3 = 1K\Omega$, the limit cycle has a shape far from a circle and it is no longer contained in a plane (blue curve).

        \section{Discussion and conclusions}

        In this work we have presented a simple and low-cost device with a rich behavior that can be modeled by a dynamical system with a Hopf bifurcation. This circuit does not require a deep understanding of electronic and can be handled by undergraduate students that have taken the basic physics courses. Notably this circuit can be seen as a physical implementation of the negative feedback three-node motif that is intensively study in biology \cite{Milo} and conceptually similar to the genetic repressilator.
    
        The performed analyses allow to understand different aspects of the proposed model. In effect, with the linear stability study we can find the the critical value of $R_3$ and estimate the frequency of the system close to the bifurcation point in an almost analytical way. From this analysis we also observe that the resistor $R_3$ plays the main role as bifurcation parameter. This is because it enters as argument of the function $u(V)$ that is the kernel of the non-linearity of this system and it is the responsible to switch between the two possible states of a node. On the other hand, resistors $R_1$ and $R_2$ have the role of controlling the characteristic times for charge and discharge of the capacitors and they are not a source of non-linearity for the system.
        
        Numerical integration of the model allows us to find the trajectories of the system in the space of voltages. We can use this information to compute the main characteristic of the signal like frequency and its shape for the whole set of possible values of the system parameters. Note that these values cannot be analytically determined since the system (\ref{equ_repressilator}) does not allow a close-form expression for its solutions.
        
        In conclusion, this system allows us to study important concepts of dynamical systems such as stability analysis and bifurcation theory. It allows also to perform analytical and numerical studies of the proposed model. Finally, we can construct a device where the theoretical results can be contrasted with the measurements.
    
        PK acknowledges financial support from SeCTyP-UNCuyo (project M028 2017-2018) and from CONICET (PIP 11220150100013), Argentina. DNR acknowledges financial support from EVC-CIN fellowship 2016, Argentina.


\begin{thebibliography}{99}
  
        \bibitem{Strogatz-book} S.H. Strogatz, Nonlinear Dynamics And Chaos: Studies in nonlinearity, (Sarat Book House, 2007).
        
        \bibitem{Perko} L. Perko, Differential Equations and Dynamical Systems, (Springer-Verlag, 2001).
        
        \bibitem{Murray} J.D. Murray, Mathematical Biology I: An Introduction, 3rd edn. Interdisciplinary Applied Mathematics, 17, (Springer, New York, 2002).

        \bibitem{HH-model} A.L. Hodgkin, and A.F. Huxley, The Journal of Physiology \textbf{116} (1952) 449-72.
        
        \bibitem{FN-model} R. FitzHugh, Biophysical Journal \textbf{1}(6) (1961) 445-466.
        
        \bibitem{BZ-model} D. Zhang, L. Gyorgyi, and W.R. Peltier, Chaos: An Interdisciplinary Journal of Nonlinear Science \textbf{3} (1993) 723-745.
    
        \bibitem{oreganator-model}  R.J. Field, and R.M. Noyes, J. Chem. Phys. \textbf{60} (1974) 1877-1884.
        
        \bibitem{brusselator-model} D. Kondepudi, and I. Prigogine, Modern Thermodynamics. From Heat Engines to Dissipative Structures. (John Wiley \& Sons, Weinheim, New York 1998).
        
        \bibitem{PP-model} R. Arditi,  and L.R. Ginzburg, Journal of Theoretical Biology \textbf{139} (1989) 311-326.
    
        \bibitem{lorentz-model} E.N. Lorenz, Journal of the Atmospheric Sciences \textbf{20} (1963) 130-141. 
        
        \bibitem{repressilator-model} M. Elowitz, and S. Leibler, Nature \textbf{403} (2000) 335-338 .
        
        \bibitem{synthetic-biology} U. Alon, An Introduction to Systems Biology: Design Principles of Biological Circuits, (CRC Press, Boca Raton, Florida, 2006).
    
        \bibitem{paper-kurst} E.H. Hellen, E. Volkov, J. Kurths, and S.K. Dana, PLoS ONE \textbf{6}(8) (2011) e23286.
        
        \bibitem{japoneses-circuitos} I.T. Tokuda, A. Wagemakers, and M.A.F. Sanju\'an, International Journal of Bifurcation and Chaos \textbf{20}(6) (2010) 1751-1760. 
        
        \bibitem{Milo} R. Milo, S. Shen-Orr, S. Itzkovitz, N. Kashtan, D. Chklovskii, and U. Alon, Science \textbf{298} (2002) 824-827. 
        
    \end{thebibliography}
\end{document}